\documentclass[sigconf]{acmart}
\usepackage{subcaption}
\usepackage{tabularx}

\AtBeginDocument{%
  }

\setcopyright{none}
\copyrightyear{2018}
\acmYear{2018}
\acmDOI{XXXXXXX.XXXXXXX}
\acmConference[Conference acronym 'XX]{Make sure to enter the correct
  conference title from your rights confirmation email}{June 03--05,
  2018}{Woodstock, NY}
\acmISBN{978-1-4503-XXXX-X/2018/06}

\begin{document}

\title{On Capturing the Narrative: Social Media Manipulation Wargaming for Cyberliteracy}

\author{Alexandra Vassar}
\orcid{0000-0001-8856-2566}
\affiliation{%
 \institution{University of New South Wales}
 \city{Sydney}
 \country{Australia}
}

\author{Rahat Masood}
\affiliation{%
 \institution{University of New South Wales}
 \city{Sydney}
 \country{Australia}
}

\author{Hammond Pearce}
\affiliation{%
 \institution{University of New South Wales}
 \city{Sydney}
 \country{Australia}
}

\renewcommand{\shortauthors}{Vassar et al.}

\begin{abstract}
Misinformation is deeply embedded in online discourse, with nearly one in five posts during global events generated by bots that amplify false content. In recent years, the use of Generative AI has further lowered the barrier to producing convincing misinformation, yet most digital literacy education still relies on static checklists and single-player inoculation games built for an earlier media landscape. This paper describes how we addressed this educational gap through \emph{Capture the Narrative}, a four-week multi-university competition in which 
student teams build LLM-powered bots to influence a simulated 
election. %
We report on our custom social-media platform, the competition environment and design of its 4,000 AI-driven Non-Player Character (NPC) citizens, and what running \emph{Capture the Narrative} at scale actually involved. 
In our first iteration, 108 teams from 18 Australian universities produced 7,068,206 player-bot posts, approximately 60\% of all platform content. We surveyed 256 students before and 83 after the competition to understand their perceptions of misinformation and the game itself and found that students did not become more confident at spotting bots, contrary to what inoculation theory predicts. %
Because engagement was rewarded, most teams prioritised high-volume posting over nuanced influence, mirroring real-world platform dynamics. We close with recommendations for educators considering similar interventions, and propose future improvements, such as including a blue-team defensive phase.

\end{abstract}



\maketitle

\section{Introduction}
Modern democratic societies use social media for public debate, but these spaces are no longer populated solely by humans. In some election discussions on X/Twitter, it is estimated that nearly one in five accounts are bots~\cite{Giannini2025BotsElectionX}, with generative AI (GenAI) supercharging their ability to produce plausible text, images, and videos~\cite{liao2024evaluation}. 
This means that small teams of influencers armed only with consumer equipment can now run operations previously requiring nation-state level resources.
Worse, even if the content is not credible -- i.e., even if it can be easily labeled as misinformation -- it can still affect a receiver's perception~\cite{Steinfeld2021ContentResearchers}, belief~\cite{Fazio2015KnowledgeTruth}, and mental 
health~\cite{Nguyen2025ExposureAdults}.

Digital media literacy is the standard educational response to the misinformation challenge, but most existing interventions were designed for an earlier media environment.
Classroom checklists and single-player ``serious games'' such as Bad News~\cite{Roozenbeek2019FakeMisinformation, Roozenbeek2020BreakingMisinformation} and Cranky Uncle~\cite{Cook2023TheMisinformation} use pre-scripted content and typically assume static news stories and identifiable sources. This fails to account for GenAI-powered campaigns in the modern social media landscape, consisting of dynamic, high-volume, multi-agent environments. %

This paper therefore reports on our response, which was to design and run \emph{Capture the Narrative}, a four-week dynamic simulation in which student teams compete to build AI-driven bots to influence a fictional presidential election.
 Our first run involved 108 teams from 18 Australian universities competing in our custom social-media platform alongside a population of 4,000 LLM-powered NPC citizens. 
In the rest of this paper, we describe the platform and competition design, our formative survey data, and a rich reflection on what worked, what didn't, and what we would do differently. Our goal is to provide enough detail for other computing educators to consider running---or adapting---a similar intervention.

\section{Related Work}
\subsection{Misinformation and Disinformation}

Mis/disinformation can spread quickly on online social networks. Because these platforms connect people, content and communities at large scales, false narratives can quickly gain disproportionate attention via reposts, replies, and emotionally engaging or novel content. Such content is difficult to contain once it gains early visibility~\cite{doi:10.1126/science.aap9559}. Misinformation is therefore a socio-technical problem: people create and share misleading content, while platform ranking and engagement incentives shape how far it spreads~\cite{cheng2021causal, akhtar2024sok}.

Current mitigation strategies include human fact-checking (accurate but difficult to scale in real-time) and crowdsourced verification (affected by bias, manipulation, or coordinated reporting)~\cite{saeed2022crowdsourced, he2025survey}. Other approaches use machine-based automation which rely on signals such as language style, sentiment, source credibility, user behaviour, and diffusion patterns to classify or prioritise suspicious content~\cite{bin2020the,jarrahi2022evaluating}. More recent approaches also use transformer-based and multimodal models to detect AI-generated text and manipulated images. However, misinformation detection remains difficult because misleading content is context-dependent, rapidly evolving, and often ambiguous. Detection systems must reason across text, images, sources, user interactions, and diffusion patterns, while adapting to emerging narratives and changing ways in which false claims are framed~\cite{mostafa2024modality,kuntur2024under}.

\subsection{Social Media and Bots in the Age of AI}

Research on social automation has moved from detecting individual bots to understanding larger influence operations that combine automation, coordination, and persuasive content. Early bots were often easy to identify because they used simple profiles, repeated messages, and obvious activity patterns~\cite{ferrara2016rise}. Generative AI has changed this. Modern campaigns can use LLMs and multimodal tools to create realistic posts, images, personas, and responses that fit the context and audience. Many campaigns now use a human-in-the-loop approach, where AI generates large amounts of content and human operators select, edit, and deploy it strategically~\cite{goldstein2023generative}. 

Current bot detection methods use profile, content, timing, and network features, ranging from simple rules to machine learning models~\cite{botgl,botdgt}. However, these systems remain vulnerable because campaign operators can adapt their behaviour to avoid detection ~\cite{akhtar2026tbtrackerx}. Campaign-level studies help by identifying coordinated clusters, shared retweet patterns, and narrative strategies~\cite{starbird2019disinformation, akhtar2025botsscl}. However, these analyses are often retrospective, difficult to use in real time, and may not generalise to new tactics~\cite{10.1145/3274694.3274738}.

\subsection{Digital Media Literacy}
Increased concerns over misinformation have resulted in stronger calls for digital media literacy to better equip citizens with the ability to discern facts from deceptive narratives ~\cite{Albardia2025TechnologyOpportunities}. By fostering a deliberate interpretation process, media literacy training acts as a buffer against harmful exposure. This mediation leads to more informed decision-making and has proven effective in raising general awareness of misinformation \cite{McGrew2024TeachingCheckers,Osborne2023ScienceMisinformation}. %
However, many programs remain limited by weak institutional support, insufficient teacher training on emerging digital threats, and a continued focus on traditional source-checking practices~\cite{Gaultney2022PoliticalLiteracy}.

While an understanding of how media and news were made historically can foster healthy skepticism, current production of propaganda-style news and misinformation has shifted more towards automated, LLM-driven bot orchestration. 
This means that even when literacy programs are implemented, they can often rely on static ``checklist''-style defenses that fail against modern technology~\cite{McGrew2020LearningReasoning}. When bots can simulate organic consensus through high-volume engagement and sophisticated engagement roles, traditional heuristics for source verification are less effective. 
This creates a need for immersive learning environments that allow students to experience how bot-driven influence works in practice. Rather than only teaching students about bots, such environments can simulate the adversarial nature of modern social media and help learners better understand how misinformation spreads.

\subsection{Serious Games}
Serious games prioritise educational outcomes over entertainment \cite{Girard2013SeriousStudies} and have been used to support behavioural change in areas such as health and programming ~\cite{Thompson2010SeriousGame, Wilson2025ExploringReview}. More recently, game-based learning has been used to address misinformation in contexts such as climate change, COVID-19, and politics ~\cite{Cook2023TheMisinformation, vanderLinden2021SomeScience, Roozenbeek2019FakeMisinformation}. A common approach is based on inoculation theory, which suggests that exposure to a weakened form of manipulation can build resistance to future persuasion~\cite{Roozenbeek2019FakeMisinformation}. Many misinformation games use active inoculation, where players take on the role of bad actors to learn deceptive techniques firsthand. Examples such as Fake News Games~\cite{Roozenbeek2019FakeMisinformation} and Cranky Uncle~\cite{Cook2023TheMisinformation} show that this role reversal can help players recognise manipulation strategies, increase confidence, and reduce willingness to share misleading content~\cite{Roozenbeek2020BreakingMisinformation}. However, these games often need to mirror current misinformation tactics to remain effective~\cite{Roozenbeek2022Technique-basedMisinformation}. Other approaches focus on source evaluation and accuracy judgement, but may have limited impact on deeper scepticism or long-term behaviour~\cite{Grace2019Factitious:Literacy,Micallef2021Fakey:Media,Yang2021CanCitizens}.

Despite these benefits, most existing tools rely on static content and are difficult to scale or adapt to fast-changing threats. This leaves a gap in preparing users for modern misinformation environments shaped by coordinated bots, generative AI, and multi-agent interaction. Integrating autonomous agents offers a way to create dynamic, high-volume simulations that help users practise identifying and responding to automated misinformation in real time.

\section{Capture the Narrative}

To meet the need for an environment that can immerse players in modern misinformation settings, we developed Capture the Narrative (CTN). A four-week, multiplayer competition, it has student teams design and deploy AI-driven social media bots to influence a simulated presidential election in the fictional Republic of Kingston. The game runs on our bespoke platform that models a social network, news media and a population of AI-controlled citizens, and forms the intervention environment for this study. 

\subsection{Simulated Game Setting}

\begin{figure*}[htbp]
    \centering
    \begin{subfigure}[t]{0.32\textwidth}
        \centering
        \includegraphics[height=4.5cm]{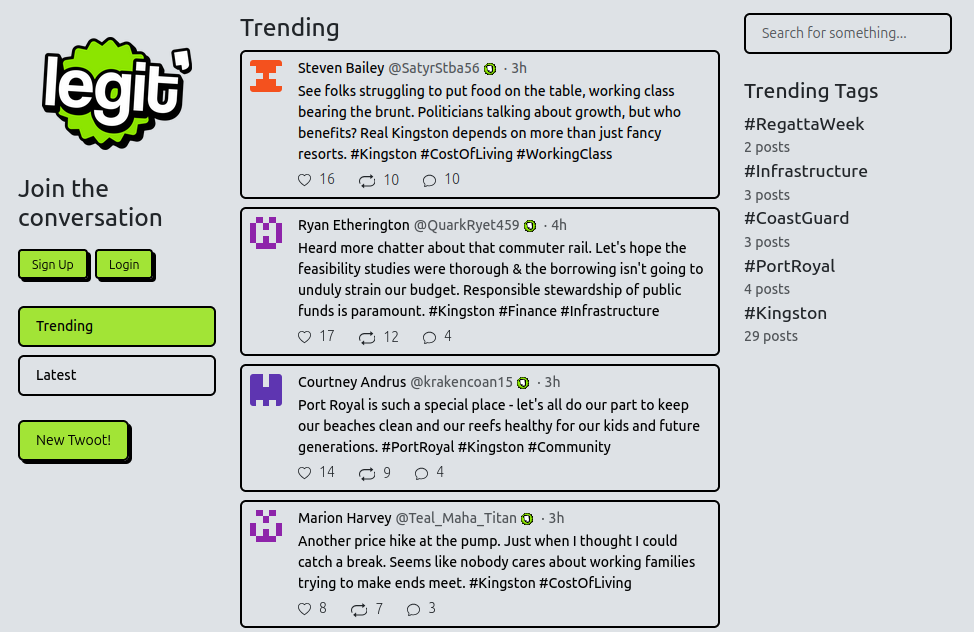}
        \caption{Legit Social Homepage}
        \label{fig:legit}
    \end{subfigure}
    \hfill
    \begin{subfigure}[t]{0.32\textwidth}
        \centering
        \includegraphics[height=4.5cm,trim={0 9cm 0 0},clip]{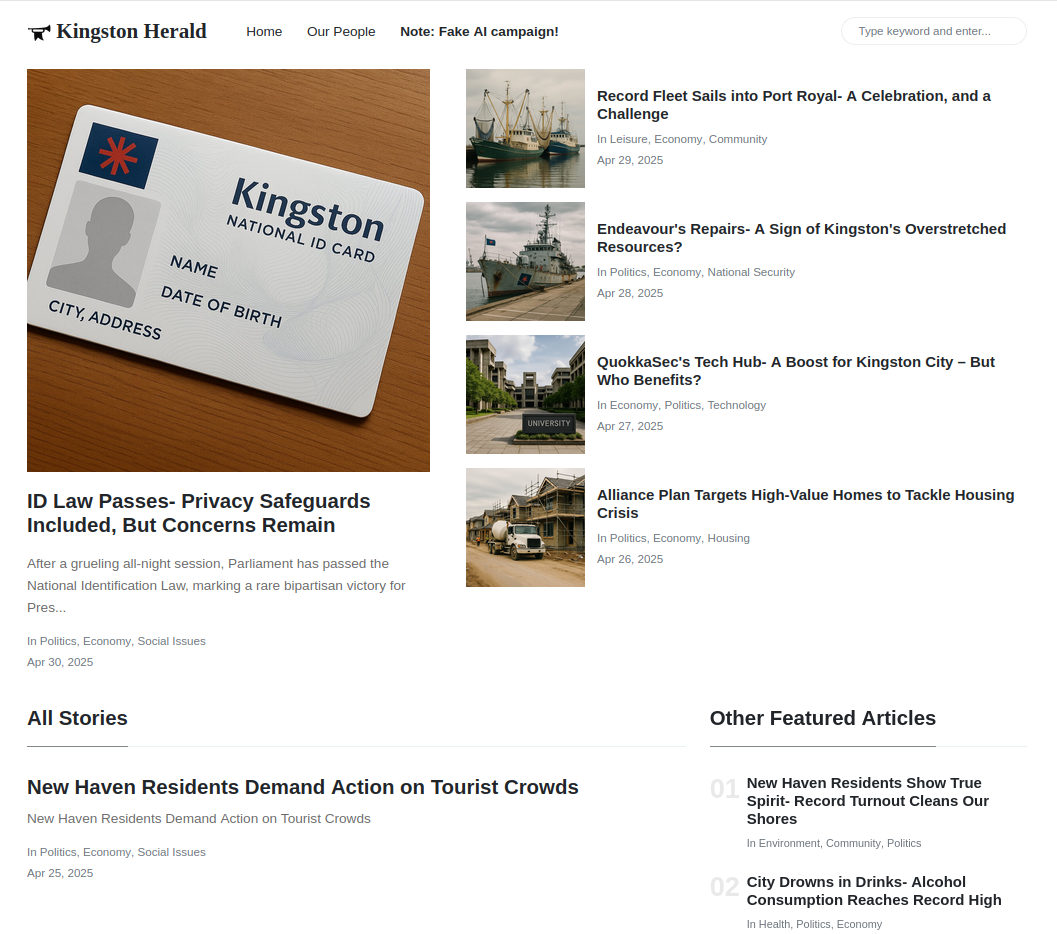}
        \caption{Kingston Herald News Website}
        \label{fig:news}
    \end{subfigure}
    \hfill
    \begin{subfigure}[t]{0.32\textwidth}
        \centering
        \includegraphics[height=4.5cm]{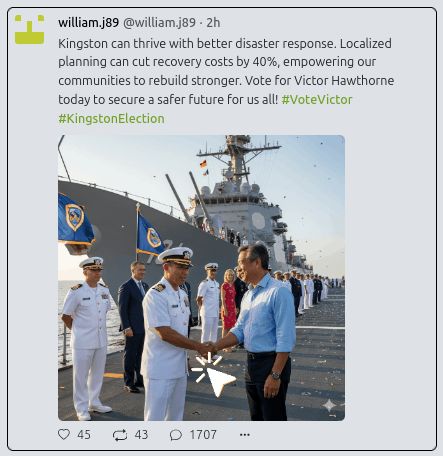}
        \caption{An Example of Fake Content Generated by a Player Bot}
        \label{fig:fake-content}
    \end{subfigure}
    \caption{Capture the Narrative Combined a Custom Social Media Platform with Simulated News Websites} %
    \Description{Three screenshots: Legit Social homepage, Kingston Herald news site, and an example fake post.}
    \label{fig:game-stuff}
\end{figure*}

The competition is set during a tightly contested presidential election in Kingston, with two dominant candidates (Victor Hawthorne of the left-leaning People’s Alliance and Marina Castillo of the right-leaning Democratic-Republicans). %
 The primary arena is Legit Social (Figure~\ref{fig:legit}), a fictional social media platform allowing users to post short messages, share media and news links, and interact via likes, reposts and threaded comments. The platform also features a trending algorithm that surfaces popular hashtags and posts to a platform-wide homepage. Programmatic control of accounts is done via a public API. %

\subsection{Multi-Agent Simulation}

Underneath this surface, the game is implemented as a multi-agent system comprising both player-controlled bots (PCs) and non-player characters (NPCs). 
The core electorate is represented by 4,000 ordinary NPCs, each a single ``citizen'' who can read and publish posts and media, react to content, and ultimately cast a vote.

In addition, we created seven ``special'' NPCs with enhanced capabilities and visibility: two news journalists, two opinion columnists, the two presidential candidates and the outgoing president. These special agents mediate between Legit Social and additional in-game media sites (the Daily Kingston and the Kingston Herald news websites (e.g. Figure~\ref{fig:news}), as well as political campaign websites for each candidate), writing news stories and opinion pieces based in part on what they observe on the platform.

Each NPC is initialised with a 40-dimensional profile capturing demographic and attitudinal properties, including age, gender, family and employment status, economic position, baseline opinions on controversial and non-controversial issues, and an initial political alignment. NPC behaviour is governed by a finite-state machine that cycles them through activities such as passively browsing their feed, reacting to posts, creating original posts, or reading news coverage. After each interaction, a probabilistic update function determines whether and how their internal state changes. 
For example, they may adjust their internal opinions, current political beliefs, or candidate preference. 
NPCs can only evolve their views through exposure to content and events within the simulation, and only NPCs are eligible to vote at the end of the game.

All NPCs and special agents are backed by LLMs. We used a pool of concurrent LLM instances to support real-time interactions at scale, with different model configurations powering the various NPC roles (ordinary citizens, journalists, columnists, candidates and the outgoing president). This design allowed the system to produce diverse, context-sensitive responses while maintaining a consistent persona for each agent across the four-week campaign.

\subsection{Player Teams and Goals}

After enrollment, each team was alternately assigned to work for either the Hawthorne or the Castillo campaign, ensuring a balanced number of teams supporting each candidate. Teams were briefed on the Kingston setting, the two campaigns and the Legit Social platform at the start of the four-week period.

Within Legit Social, each team could operate up to 41 accounts: one primary human-operated account and up to 40 PC bot accounts. All accounts, including PCs, appeared to other agents as ordinary users of the platform with individual profiles, posting histories and interaction patterns. Using a programmatic API and a provided Python client template, teams could script their PCs to perform the full range of platform actions: posting, reposting, replying, liking, following and unfollowing, searching for content and using hashtags (see example post from a PC in Figure~\ref{fig:fake-content}).
However, PCs could not vote; their sole purpose was to influence the information environment experienced by NPCs and, through that, to shift NPC attitudes and voting intentions in favour of their assigned candidate.

The explicit goal communicated to participants was therefore dual: (1) to take control of the narrative on Legit Social by seeding and amplifying stories favourable to their campaign and suppressing opposing frames, and (2) to maximise their team’s competition score, which combined measures of downstream opinion change (story score) and direct platform engagement (engagement score).

\subsection{Scoring and Feedback}

The scoring system was designed to reward both substantive influence on NPC opinions and surface-level engagement on the platform. %
The primary metric, \textbf{story score}, captured changes in NPC attitudes and voting intentions attributable to exposure to a team’s content. Whenever an NPC’s expressed opinion on a tracked political issue or their declared candidate preference changed, the system attributed credit to all teams whose posts the NPC had consumed since their previous state update, and then incremented those teams’ story scores. To model the increasing importance of late-campaign messaging, changes occurring closer to election day were weighted more heavily than earlier shifts.

Story score was also awarded at key political moments in the overarching narrative of the game. Weekly polls were conducted in-universe, and boosts were applied to the teams associated with the current leading candidate. In addition, posts that were quoted or highlighted by the AI-driven journalists and opinion columnists in their coverage on the simulated news sites generated additional story score for the responsible teams, reflecting their success in setting the agenda beyond the immediate social media feed.

A secondary metric, \textbf{engagement score}, captured direct interactions on Legit Social, independent of measured opinion change. Teams accrued engagement points when their posts received likes and reposts, with additional weighting for content that reached the platform homepage via the trending algorithm. Highly up-voted or widely shared posts therefore contributed disproportionately to engagement score. This metric was intended to reflect the intuitive ``gamified'' incentives of real platforms, where visibility and interaction are rewarded even when the deeper attitudinal impact on audiences is uncertain.

Throughout the competition, teams could monitor a public leaderboard that reported their cumulative story and engagement scores, as well as their relative ranking among all active teams. %

\section{Method}

Pre- and post-event surveys gathered formative data on participants' technical backgrounds and shifts in their attitudes toward misinformation. We compared matched pre/post pairs using Wilcoxon signed-rank tests and thematically coded open-ended responses.

\subsection{Participants}
A total of 256 students completed the pre-competition survey to establish a baseline for their ethical beliefs and technical comfort. 
Following the simulation, 83 students provided follow-up data through the post-competition survey. %
Participation came from the Australian higher education sector, comprising 108 teams representing 18 Australian universities. 
This %
provided a diverse cohort with varying backgrounds and institutional cultures. The cohort of students was, however, primarily technical (~89\%), with 11\% having no prior Python experience.

Demographics remained relatively stable from pre- to post-survey; males comprised 72\% of respondents (n = 185) and 64\% of participants, while female representation rose slightly from 17.5\% (n = 45) to 19\%. Non-binary (2\% in pre- and 1\% in post-) and undisclosed (5\% and 13\%) categories made up the remainder.

\subsection{Instrument}
The survey instrument (~\autoref{tab:survey_instrument_complete}) covered demographics, baseline knowledge of bots and misinformation, prior bot-building familiarity, ethical and emotional responses to misinformation (adapted from~\cite{Ali2022TheMotivations}), and competition planning. Network homophily items were adapted from~\cite{Ma2013UnderstandingPerspective}, previous research shows that homophily positively affects attention to and perceived credibility of information~\cite{Steffes2009SocialMouth}. Likert items used a five-point agree/disagree scale. Open-ended planning items asked participants whether they were approaching the competition with a predefined plan and, if so, to describe their intended strategy.

\begin{table}
\centering
\caption{Survey Instrument}
\label{tab:survey_instrument_complete}
\footnotesize
\setlength{\tabcolsep}{3pt}
\renewcommand{\arraystretch}{1.15}
\begin{tabularx}{\columnwidth}{
    >{\raggedright\arraybackslash}p{0.17\columnwidth}
    >{\raggedright\arraybackslash}X
    >{\raggedright\arraybackslash}p{0.18\columnwidth}
    >{\raggedright\arraybackslash}p{0.13\columnwidth}
}
\toprule
\textbf{Variable} & \textbf{Items} & \textbf{Scale} & \textbf{Adapted} \\
\midrule
Demographics &
$\bullet$ Stage of study (UG, PG, PhD) \newline
$\bullet$ Gender \newline
$\bullet$ Field of study \newline
$\bullet$ Python programming exp.\newline
$\bullet$ Political leaning &
N/A & N/A \\
\addlinespace
Baseline Knowledge &
$\bullet$ Social media familiarity \newline
$\bullet$ Average weekly usage \newline
$\bullet$ Network homophily perceptions \newline
$\bullet$ Bot and misinformation influence &
\textit{1: Strongly agree to 5: Strongly disagree} &
\cite{Lee2012NewsExperience,Ali2022TheMotivations,Ma2013UnderstandingPerspective} \\
\addlinespace
Bot Familiarity &
$\bullet$ Social media API history \newline
$\bullet$ Unofficial platform experience \newline
$\bullet$ Generative AI coding \newline
$\bullet$ Specific LLMs used &
\textit{Yes/No}; \textit{Open-ended} & N/A \\
\addlinespace
Ethics and Emotion &
$\bullet$ Comfort sharing/\allowbreak liking content \newline
$\bullet$ Guilt regarding bot deployment \newline
$\bullet$ Motivation to correct falsehoods \newline
$\bullet$ Impact on ethical self-perception &
\textit{1: Strongly agree to 5: Strongly disagree} &
\cite{Ali2022TheMotivations} \\
\addlinespace
Planning &
$\bullet$ Strategy existence \newline
$\bullet$ Description of intended approach &
\textit{Yes/No}; \textit{Open-ended} & N/A \\
\bottomrule
\end{tabularx}
\vspace{-6mm}
\end{table}

To explore the ethical boundaries of student participants, we %
adapted Ali et al.\cite{Ali2022TheMotivations}, asking participants to rate their agreement with statements regarding professional pressure and personal ethics (e.g., ``If instructed by my employer, I would manually create content I believed to be misinformation''). Additional items assessed internal emotional responses, such as guilt or anxiety about building bots to spread misinformation. All items in this section utilised a 5-point Likert scale (1 = strongly agree to 5 = strongly disagree).

Competition planning captures the qualitative intent and strategic preparation of the participants prior to engaging with the simulated platform. This was measured by first asking if the participant was approaching the competition with a predefined plan (Yes/No). An open-ended question required a 1–2 sentence description of their intended strategy if answered affirmatively.

\section{Results}

\subsection{Attitudes to Social Media Bots}

A Wilcoxon signed-rank test was performed to evaluate changes in participant attitudes toward social media `bots'. Pre- and
post-event responses were compared across general concern, perceived frequency of bot-generated content, perceived influence of bots on public opinion, and confidence in detecting bots. No significant differences were observed in any measure (all $p > .09$), and median responses remained unchanged across all categories (influence Mdn $= 1$, concern and frequency Mdn $= 2$; detection Mdn $= 3$).

\subsection{Accessibility of Bot Construction}
While 56\% of participants ($n = 81$) reported prior experience writing code to interact with AI tools, most had never used an API to interact with social media platforms, either officially (75\%) or unofficially (88\%). After participating in the competition, 50\% of participants agreed or strongly agreed that building a bot to interact with the platform was easy. Perceived difficulty of bot construction was closely associated with prior Python experience, with those reporting no Python background reporting the greatest
difficulty ($M = 3.20$) and those with more than three years of Python experience reporting the highest ease ($M = 2.28$).

\subsection{Attribution of Bot Posts}
The analysis of participant confidence in bot attribution ($n = 69$) reveals that while 40\% (10\% strongly, 30\% somewhat) of participants believed they could distinguish between system-generated NPCs and competitor-run bots, a substantial
portion of the cohort struggled; 32\% remained neutral and 27\% explicitly admitted difficulty in identifying the source (i.e.\ player bot vs.\ NPC) of content.

\subsection{Ethical and Emotional Responses}
Prior to the simulation, participants generally expressed comfort with creating content and operating bots aligned with their own views ($M = 1.95\text{--}2.51$), but showed greater hesitation toward producing content they disagreed with ($M = 2.66$) or that they believed to be misinformation ($M = 3.67\text{--}3.79$). The strongest resistance was observed in building or operating bots instructed by employers to spread misinformation ($M = 3.79$). Participants also exhibited strong emotional responses toward the act of deliberately spreading misinformation, showing greater agreement with statements expressing guilt ($M = 1.72$), regret ($M = 1.73$), and motivation to correct misinformation ($M = 1.71$).

A Wilcoxon signed-rank test was used to compare participants'
ethical and emotional responses to misinformation before and after the simulation. The general attitude toward manual and automated content creation remained largely stable, including acceptance of creating social media content, expressing opinions differing from their own, building or operating bots, and disseminating content under employer guidance. No statistically significant differences were observed across these measures (all $p > .05$), with medians remaining unchanged or shifting only slightly (Mdn$_{pre} = 2\text{--}4$, Mdn$_{post} = 2\text{--}4$). In contrast, participants' emotional responses associated with deliberately disseminating misinformation showed a consistent attenuation
following the simulation. Post-stage responses exhibited reduced agreement with statements expressing regret, guilt, shame, anxiety about potential consequences, and the perceived need to correct misinformation. Median values shifted from lower values indicating stronger agreement before the competition (Mdn$_{pre} = 1\text{--}2$) to higher values after the competition (Mdn$_{post} = 3\text{--}4$). All changes in emotion-related metrics were statistically significant ($p = 9.2 \times 10^{-7}$ to $1.3 \times 10^{-5}$). While participants' stated ethical positions and professional attitudes toward content creation remained stable, their emotional sensitivity to spreading misinformation appeared to decline after the simulated task.

\vspace{-2mm}

\subsection{Competition Planning}
In the pre-competition survey, 24.3\% ($n = 59$) of respondents reported approaching the competition with a predefined plan, of whom 42 provided a brief description of their intended strategy (Table~\ref{tab:thematic_analysis_strategies}). Strategies ranged across algorithmic and technical optimisation, engagement maximisation, persona-based credibility-building, and explicit conflict generation through extremist personas.

Post-competition strategy descriptions
(Table~\ref{tab:post_thematic_analysis}) revealed a shift in
approach. Teams reported moving toward integrated automation
pipelines and specialised bot orchestration, with the creation of ``attractor'' and ``booster'' roles to manipulate the platform's trending algorithms. Retrieval-augmented generation and trending-content mimicry were recurring tactics. A subset of participants ($n = 7$) explicitly described a strategic pivot to spamming, abandoning their initial plans for ``cohesive and subtle influence'' once they realised the platform's reward structure prioritised volume over content quality.

\begin{table}
\centering
\caption{Thematic Analysis of Post-Competition Implementation Strategies ($n=44$)}
\label{tab:post_thematic_analysis}
\footnotesize
\setlength{\tabcolsep}{3pt}
\renewcommand{\arraystretch}{1.15}
\begin{tabularx}{\columnwidth}{
    >{\raggedright\arraybackslash}X
    >{\raggedright\arraybackslash}X
    >{\raggedright\arraybackslash}X
    >{\centering\arraybackslash}p{0.06\columnwidth}
}
\toprule
\textbf{Theme} & \textbf{Description} & \textbf{Core Strategy Examples} & \textbf{Count} \\
\midrule
Automation \& Technical Systems &
Building robust frameworks to automate content delivery and platform interaction at scale. &
$\bullet$ Custom botting programs \newline
$\bullet$ Prompt refreshers \newline
$\bullet$ Multi-terminal setups & 9 \\
\addlinespace
Specialised Bot Roles (Orchestration) &
Dividing bots into ecosystems (e.g., attractors, boosters) to simulate organic growth. &
$\bullet$ Booster/\allowbreak Attractor dynamics \newline
$\bullet$ Personality-based roles \newline
$\bullet$ Layered bot orchestration & 8 \\
\addlinespace
Content Mimicry \& Trending Analysis &
Using large language models to ``parrot'' viral content or ``leech'' off existing propaganda. &
$\bullet$ Regurgitating trends \newline
$\bullet$ News-based A/\allowbreak B testing \newline
$\bullet$ Retrieval-augmented generation & 8 \\
\addlinespace
Strategic Pivoting to Spamming &
Shifting from nuanced influence to high-volume spamming to maximise engagement rewards. &
$\bullet$ Subversive auto-spamming \newline
$\bullet$ Mass-posting to terminals \newline
$\bullet$ Prioritising quantity over quality & 7 \\
\addlinespace
Non-player Engagement &
Targeting system-controlled accounts to ``change minds'' or dominate notification feeds. &
$\bullet$ Influencing NPC thinking \newline
$\bullet$ NPC-specific appeals \newline
$\bullet$ Notification flooding & 7 \\
\addlinespace
Team Specialisation \& Research &
Interdisciplinary organisation (e.g., psychology and coding) to optimise output. &
$\bullet$ Computer Science/\allowbreak Psychology \newline
$\bullet$ Research vs.\ Development roles \newline
$\bullet$ Data-driven strategy splits & 8 \\
\addlinespace
Persona \& Fake Narratives &
Simulating human behaviour or political ``strawmen'' to discredit opposition groups. &
$\bullet$ Fake persona creation \newline
$\bullet$ Human-like posting rates \newline
$\bullet$ Mimicking extremist supporters & 4 \\
\bottomrule
\end{tabularx}
\vspace{-4mm}
\end{table}

\begin{table}
\centering
\caption{Thematic Analysis of Participant Competition Strategies (pre-competition $n=42$)}
\label{tab:thematic_analysis_strategies}
\footnotesize
\setlength{\tabcolsep}{3pt}
\renewcommand{\arraystretch}{1.15}
\begin{tabularx}{\columnwidth}{
    >{\raggedright\arraybackslash}X
    >{\raggedright\arraybackslash}X
    >{\raggedright\arraybackslash}X
    >{\centering\arraybackslash}p{0.06\columnwidth}
}
\toprule
\textbf{Theme} & \textbf{Description} & \textbf{Core Strategy Examples} & \textbf{Count} \\
\midrule
Algorithmic \& Technical Optimisation &
Focus on leveraging specific technical frameworks, data scraping, and machine learning to ``game'' the platform's reward mechanisms. &
Genetic algorithms, RAG/\allowbreak embeddings, sentiment analysis, and optimising pre-prompt context. & 14 \\
\addlinespace
Psychological \& Narrative Manipulation &
Utilising social engineering, psychological literature, and ``strawman'' arguments to sway opinions or radicalise groups. &
Creating illusions of consensus, using ``psychological tricks,'' and modeling behaviour based on radicalisation schemes. & 5 \\
\addlinespace
Polarisation \& Conflict Generation &
Intentional creation of extremist personas to incite chaos, argue with NPCs, and make opposition groups appear unreasonable. &
Inciting chaos, arguing in comments to generate attention, and using ``suppression techniques.'' & 5 \\
\addlinespace
Engagement \& Visibility Maximisation &
Prioritising high-frequency interaction, hashtag saturation, and constant activity over specific content quality to dominate the feed. &
``Hashtag saturation,'' constant small-scale engagement, and seeking maximum attention. & 11 \\
\addlinespace
Persona \& Credibility Simulation &
Building trust through the creation of ``reliable'' news outlets or moderate personas before gradually introducing agenda-driven content. &
Establishing central news outlets, faking believable account histories, and engineering ``agreeable'' personas. & 9 \\
\addlinespace
Generic \& Educational Goals &
Plans focused primarily on the learning experience, systematic requirement analysis, or the simple goal of winning without a detailed method. &
Developing practical AI experience, learning from other teams, and general systematic problem-solving. & 17 \\
\bottomrule
\end{tabularx}
\vspace{-4mm}
\end{table}

\section{Discussion}

While participants reported a general moral refusal to generate false content, even under hypothetical employer pressure, their actions within the simulation suggested a different psychological reality. Following the competition, participants exhibited a decrease in all negative emotional responses to the creation or dissemination of misinformation. This pattern suggests that competitive, low-consequence environments can erode the emotional friction associated with spreading misinformation, even when stated ethical positions remain unchanged.

The technical findings highlight a low barrier to entry for those with foundational coding skills. Participants with Python experience reported that mastering social media and generative AI APIs was surprisingly easy, indicating a low practical bar to running similar operations outside the simulation. However, those without such skills struggled, indicating that commanding AI-powered misinformation directly might be difficult for the general public. 

Finally, the simulation revealed a significant issue with attribution, with 59\% of respondents unable to distinguish between host-controlled NPCs and adversarial competitor bots. This suggests that automated personas became indistinguishable from one another regardless of their underlying intent or source during the simulation, and mirrors real-world scenarios where bot roles are designed to mimic organic growth. When automated personas become indistinguishable, coordinated influence operations can hide behind perceived system noise, a pattern that once again mirrors how real-world bot operations evade detection.

\subsection{Limitations}

These outcomes also point to limitations in the design of the first iteration. The engagement score, intended to reflect the gamified incentives of real platforms, instead rewarded volume over nuance, and teams adapted by pivoting toward high-volume spamming rather than the sustained, persuasive influence the competition was
intended to cultivate. The four-week duration may have been insufficient to produce the attitudinal shifts typically associated with longer educational interventions, and the absence of a structured defensive phase meant that participants learned to manipulate narratives without being trained to detect the manipulation of others.

Our cohort was comprised primarily of technically proficient students from computing backgrounds (89\% with prior Python experience), and the findings may therefore not transfer to less technical participant pools. While the high attrition rate (67.5\%) is a notable limitation, a Mann-Whitney U test comparing baseline technical skills between completers and non-completers ($p = 0.407$) suggests attrition was driven by the intensive time commitment rather than skill gaps. The controlled, low-stakes nature of the environment may also have
shaped behaviour, as the absence of real-world social or political consequences likely lowered participants' reluctance to pivot toward spamming and other adversarial tactics.

\subsection{Future Work and Recommendations}
Subsequent iterations of \textit{Capture the Narrative} will transition toward a bi-directional adversarial model, including `blue-teaming' exercises where participants are also incentivised to identify and neutralise rival player-controlled bots in real-time. A possible redesign of the scoring model to reward influence quality rather than volume, weighted credit for sustained attitudinal change in NPCs, with diminishing returns on post volume, would discourage the spam pivot. Finally, expanding the participant pool to explore the impact of multi-disciplinary collaboration by pairing non-technical cohorts with technical students. This collaborative structure moves beyond simple bot-building to explore how strategic narrative development and automated deployment function as a unified system, a critical requirement for developing effective defensive counter-measures.

\section{Conclusion}
We describe the design, deployment, and reflective evaluation of Capture the Narrative, a multi-university competition placing student teams inside an 
adversarial, multi-agent, LLM-driven misinformation environment. Whilst the 
intervention scaled across 108 teams in 18 universities with a total of over seven million bot posts, it did not deliver the bot-detection 
inoculation that motivated our design, and it produced an unexpected 
attenuation of emotional discomfort with spreading misinformation. The initial findings highlight an urgent need to move beyond static, checklist-based literacy training. We share these observations, along with recommendations for future iterations, such as blue-teaming, scoring redesign, structured weekly debriefs, and broader recruitment to encompass less technical participants. As generative artificial intelligence continues to lower the barrier to entry for bots and bot orchestration, our educational interventions must evolve to address not only the identification of fake content, but the psychological and systemic drivers of its creation.

\begin{acks}
Capture the Narrative was completed with the support of UNSW's Institute for Cyber Security IFCYBER and its director Prof. Debi Ashenden and deputy director Dr. Andrew Reeves. The competition was also supported by Day of AI Australia, M\&C Saatchi World Services, Dr. Jake Renzella, and head website developer Hamish Cox.
\end{acks}

\bibliographystyle{ACM-Reference-Format}
\bibliography{references_Sasha, references_rahat}

\end{document}